\documentstyle[12pt]{article}
\begin{document}
\centerline{\bf Investigations on a bi-chiral boson}
\bigskip\bigskip
\centerline {Anisur Rahaman}
\smallskip
\centerline{Saha Institute of Nuclear Physics,Block AF,
Bidhannagar, Calcutta 700 064, INDIA}

\smallskip
\centerline{At present}
 \centerline {Hooghly Mohsin College, Chinsurah,
Hooghly-712101, INDIA}
\bigskip\bigskip\bigskip
 \centerline{\bf Abstract}
\smallskip
Chiral  boson fields in (1+1) dimensions obey a first order equation of
motion. A second order equation which classically
describes two chiral boson fields of the same chirality leads, however,
to an unacceptable quantum theory.

\vfill\eject
\bigskip
A real massless boson field satisfies the equation
\begin{equation}
\Box\phi=0.\end{equation} in (1+1) dimensions, this means that the
product of the two lightcone components of the momentum vanishes.
There are two ways of solving this equation, and one may regard
the massless boson field as being the assembly of two {\it chiral}
boson fields, one with the + component of the momentum vanishing,
and the other with the - component vanishing. Thus a chiral boson
field satisfies an equation
\begin{equation}
\partial_{\pm}\phi  =   0,
\end{equation}
where
\begin{equation}
\partial_+ = \partial_0 + \partial_1.\end{equation}
The  equation  of motion is a first order one, and this makes the
construction   of   a   quantum   theory   somewhat   nontrivial.
Siegel$^{\cite{WS}}$  started  with the Lagrangian for a massless
scalar  field  and  tried  to  make  it  chiral  by  imposing   a
constraint.  A  term  of  degree  three  involving  the square of
$\partial_\pm\phi$ and an auxiliary field had to  be  introduced.
Subsequently,  Floreanini  and  Jackiw$^{\cite{FJ}}$ discovered a
quadratic Lagrangian  involving  no  auxiliary  field,  but  this
Lagrangian  does  not  possess  explicit  Lorentz  invariance. Of
course, the {\it theory} is Lorentz invariant$^{\cite{FJ}}$.

If  a  lightcone  component  of the momentum vanishes, its square
vanishes too, so one might be tempted  to  think  that  a  second
order equation would also describe a chiral boson field. But the second
order  equation  is  a weaker restriction on the field, and it is
more natural to expect something similar to a full  field  degree
of freedom rather than the half degree that a chiral boson field is. As
the boson field is definitely chiral, one is then led to expect roughly
a  pair  of chiral boson fields of the same chirality, in contrast to a
scalar field which is equivalent to two chiral boson fields of opposite
chiralities.  In  this  note  we  analyse  in  detail  the theory
underlying the second  order  equation.  The surprising result is
that one does have two chiral boson fields but in quite a   complicated
way,  and  in  fact  the corresponding quantum theory is not well
defined.

We  start with the lagrangian   density
\begin{equation}
{\cal   L}   =  {1\over2}(\partial_+\phi)^2.
\end{equation}
This Lagrangian density is not manifestly Lorentz
invariant,  but  Lorentz  invariance   is   maintained, as is clear
from the Euler - Lagrange  equation
\begin{equation}
\partial_+^2\phi  =   0.
\label{1}
\end{equation}

The components of the energy momentum tensor $T_{\mu\nu}$ are given by

\begin{equation}
T_{00} = {1\over 2}(\dot\phi^2 - \phi'^2),
\end{equation}
\begin{equation}
T_{01} = \dot\phi\phi' + \phi'^2,
\end{equation}
\begin{equation}
T_{11} = {1\over 2}(\dot\phi^2 - \phi'^2),
\end{equation}
\begin{equation}
T_{10} = - \dot\phi\phi' - \dot\phi^2,
\end{equation}

and the conservation of $T_{\mu\nu}$ is obvious.

The equation (\ref{1}) requires $\phi$ to be of the form
\begin{equation} {\phi} ={\phi_0(x^-)} + {x^+\over 2}{\phi_1(x^-)},
\end{equation}
where
\begin{equation}
x^\pm = x^0\pm x^1.\end{equation}
This  indicates  the  existence  of two chiral boson fields of the same
chirality, {\it i.e.}, travelling in the same direction (right).

The  momentum corresponding  to the field ${\phi}$ is given by
\begin{equation}
\pi = {\partial{\cal L}\over{\partial{\dot\phi}}}=\partial_+\phi
= \phi_1(x^-)
\end{equation}
The  Hamiltonian  density is obtained by the Legendre
transformation
\begin{equation}
{\cal H} = \pi{\dot\phi} - {\cal L}
\end{equation}
and the total Hamiltonian is found to be
\begin{equation}    H
=\int dx [-\phi_1\phi_0'+{1\over 4}\phi_1^2],
\label{2}
\end{equation}
where an integration by parts has been carried out: the field is
assumed to vanish sufficiently fast at spatial infinity.
This   Hamiltonian  is  not  hermitian.  To  get  a  hermitian
Hamiltonian one must take
\begin{equation} H
=\int dx [-{1\over 2}\phi_1\phi_0'-{1\over 2}\phi_0'\phi_1
+{1\over 4}\phi_1^2].
\end{equation}

The Fourier expansion of $\phi$ may be taken as
\begin{equation}
\phi_0    ={1\over{\sqrt{2\pi}}}   \int_{-\infty}^{+\infty}{{dk}\over
{|k|}}a(k) e^{ikx^-},
\end{equation}
\begin{equation}
\phi_1    ={i\over{\sqrt{2\pi}}}   \int_{-\infty}^{+\infty}{dk}
{|k|}b(k) e^{ikx^-},
\end{equation}
where the hermiticity of $\phi$ requires that
\begin{equation} {a^{\dag}(k)} = a(-k), {b^{\dag}(k)} = -b(-k)
\end{equation}

The quantization condition is
\begin{equation}
[\phi(x) , \pi(y)]_{x^0=y^0} = i\delta (x^1 - y^1),
\end{equation}
which is equivalent to
\begin{equation}
[\phi_0(x^-),\phi_1(y^-)] = i\delta(x^- - y^-),
\end{equation}
and hence to
\begin{equation}
[a(k) , b(l)] = \delta(k + l).
\end{equation}
All other commutators are zero.
Now we  define an annihilation operator and a creation
operator for a real parameter $\lambda$ by
\begin{equation}
c(k) = {\lambda^{-1}a(k) - \lambda |k|b(k)\over\sqrt {2|k|}},
c^{\dag}(k) = {\lambda^{-1}a(-k) + \lambda |k|b(-k)\over\sqrt{2|k|}}.
\end{equation}
The commutation relations satisfied by $c(k)$  and  $c^{\dag}(k)$
are
\begin{equation}
[c(k) , c^{\dag}(l)] = \delta (k-l),~
[c(k), c(l)] = 0.
\end{equation}
Substituting the values of $\phi_0$ and $\phi_1$ in the expression
of $H$, one obtains
\begin{equation}
H = \int_{-\infty}^{+\infty} dk~{1\over 2}[ka(-k)b(k) + kb(k)a(-k)
-{1\over 2}k^2b(k)b(-k)],
\end{equation}
which can also be written as
\begin{eqnarray}
H     &=&    -{1\over    2}    \int_{-\infty}^{+\infty}    dk~k
[c^{\dag}(k)c(k) +
c(k)c^{\dag}(k)] +
{1\over 8\lambda^{2}}\int_{-\infty}^{+\infty} dk~|k|
[c^{\dag}(k)c(k)\nonumber\\&+&c(k)
c^{\dag}(k)-c(k)c(-k)-c^{\dag}(-k)c^{\dag}(k)].
\end{eqnarray}
The energy of the system depends on the modes k that
are excited. As k runs from $-\infty
$  to  $+\infty$  the  energy has no lower (or upper)  bound,  as
may be seen by making $\lambda$ large. So a
vacuum cannot be defined by the condition $c(k)|0>=0$.

To compair with Floreanini Jackiw form of Lagrangian for right moving
chiral boson field let us start with Siegel action for right moving chiral
boson field:
\begin{equation}
S = {1\over 2}\int d^2x[\partial_+\phi\partial_-\phi +
\lambda(\partial_+\phi)^2]. \label{3}
\end{equation}
Starting from the action (\ref{3}) a Lagrangian density for right moving
chiral bosonfield can be derived where Lorentz invariance is not manifested
$^{\cite{FJ}}$:
\begin{equation}
{\cal L} = -\dot\phi\phi' - \phi'^2.
\label{4}
\end{equation}
So it is sufficient if it can be shown that the energy corresponding to the
Lagrangian density (\ref{4}) is bounded. The canonical momenta corresponding
to the field $\phi$ is
\begin{equation}
\pi_\phi = -\phi'.
\end{equation}
$\pi_\phi + \phi' = 0$ is a second class constraint itself. So Dirac's
prescription for quantization of second class constrained system has to be
followed. The dirac bracket is found out to be
\begin{equation}
[\phi(x),\phi(y)]_{Dirac} = {1\over 4}\epsilon(x - y).
\label{5}
\end{equation}
The reduced Hamiltonian is
\begin{equation}
H_r = \int dx  \phi'^2.
\label{6}
\end{equation}
Using equation (\ref{5}) and (\ref{6}) one can show that the field
$\phi$ satisfy the equation
\begin{equation}
\partial_+\phi = 0.
\end{equation}
Since the Hamiltonian (\ref{6}) is possitive definite the energy of this
system is bounded where as the energy of the system described by the
Hamiltonian (\ref{2}) is not bounded. Thus the system of bichiral bosons,
although classically interesting, does not correspond to any quantum field
theory.

I wish to acknowledge the financial support provided by CSIR.

\end{document}